\documentclass[aps,prd,superscriptaddress,nofootinbib,twocolumn]{revtex4-1}

\usepackage{amsmath,amssymb,amsbsy,amstext,amsfonts,mathrsfs,latexsym,bm}
\usepackage[colorlinks=true,urlcolor=blue,anchorcolor=blue,citecolor=blue,linkcolor=blue,pagecolor=blue,linktocpage=true,pdfa=true]{hyperref}
\usepackage{graphicx}

\begin{document}

\title{Stringy black-hole gas in $\alpha'$-corrected dilaton gravity}

\author{Jerome Quintin}
\email{jquintin@physics.mcgill.ca}
\thanks{Vanier Canada Graduate Scholar}
\affiliation{Department of Physics, McGill University, Montr\'eal, Qu\'ebec H3A 2T8, Canada}

\author{Robert H.~Brandenberger}
\email{rhb@hep.physics.mcgill.ca}
\affiliation{Department of Physics, McGill University, Montr\'eal, Qu\'ebec H3A 2T8, Canada}

\author{Maurizio Gasperini}
\email{gasperini@ba.infn.it}
\affiliation{Dipartimento di Fisica, Universit\`a di Bari, Via G.\ Amendola 173, 70126 Bari, Italy}
\affiliation{Istituto Nazionale di Fisica Nucleare, Sezione di Bari, Via E.\ Orabona 4, 70125 Bari, Italy}

\author{Gabriele Veneziano}
\email{Gabriele.Veneziano@cern.ch}
\affiliation{Coll\`ege de France, 11 place M.\ Berthelot, 75005 Paris, France}
\affiliation{Theory Department, CERN, CH-1211 Geneva 23, Switzerland}

\begin{abstract}
We discuss the properties of the gas of primordial `stringy' black holes possibly formed in the high-curvature phase preceding the bouncing transition to the phase of standard cosmological evolution. We show that the regime dominated by such a string-hole gas can be consistently described by explicit solutions of the string effective action including first-order $\alpha'$ corrections. We present a phase space analysis of the stability of such solutions comparing the results obtained from different actions and including the possibility of $O(d,d)$-symmetric configurations.
\end{abstract}

\maketitle

\section{Introduction}

Since the rise of string theory as an effort to unify quantum field theory and general relativity, there has
been a number of attempts to construct very early Universe cosmological scenarios embedded in string theory.
Notable string cosmologies include string gas cosmology \cite{Brandenberger:1988aj,Brandenberger:2011et},
pre-Big Bang cosmology \cite{Gasperini:1992em,Gasperini:1996fu,Veneziano:2000pz,Gasperini:2002bn,Gasperini:2007vw}
(see also the review \cite{Lidsey:1999mc}), and
Ekpyrotic cosmology \cite{Ekpyroticrefs,Lehners:2008vx}. There has also been a lot of effort put
into trying to build a stringy realization of inflationary cosmology (see, e.g.,
Refs.~\cite{InflationinST,Silverstein:2008sg,Cai:2014vua}),
though with limited success, given the difficulty of finding (quasi-)de Sitter solutions in the string landscape
(see, e.g., Refs.~\cite{Dasgupta:2014pma,Kutasov:2015eba} and also \cite{dSinSTprob1} and \cite{dSinSTprob2}).
Overall, current string cosmologies have led to interesting predictions, but the theories often remain
incomplete, or conceptual issues persist.
Nevertheless, studying string cosmology might be one of the best approaches to test the validity of string theory.

A common feature of many string cosmologies is that they do not start with an initial Big Bang singularity.
In string gas cosmology and pre-Big Bang cosmology, it is the T-duality of string theory that protects the
models from reaching a singularity. T-duality roughly states that a small value of the `radius of the Universe'
($R$) is equivalent to a large value of the radius. More precisely, the symmetry goes as $R\rightarrow\alpha'/R$,
where $\alpha'\sim\ell_\mathrm{s}^2$ is the string theory dimensionful parameter related to the fundamental string
length $\ell_\mathrm{s}$. Thus, one expects $R\sim\ell_\mathrm{s}$ to define a minimal length scale at which
point the Universe experiences a curvature bounce, i.e., a transition from growing to decreasing spacetime curvature. Details of
how this is realized dynamically remains a challenge, but there has been recent progress
in the context of string gas cosmology \cite{StringGasinDFT}. In pre-Big Bang
cosmology, the duality is called the scale factor duality \cite{Tseytlinrefs,Veneziano:1991ek,Sen:1991zi},
and the symmetry goes as $a \rightarrow 1/a$, where $a$ is the scale factor. Again, resolving the singularity
dynamically in this context is nontrivial but can be realized, for instance, with a nonlocal potential
\cite{Gasperini:1992em,Gasperini:1996np}, with quantum loop corrections \cite{Brustein:1997cv,Cartier:1999vk,Tsujikawa:2002qc},
or with limiting curvature \cite{LimitingCurvature} (see also the reviews
\cite{Gasperini:2002bn,Gasperini:2007ar,Gasperini:2007zz}), though the latter might be unstable
to cosmological perturbations \cite{Yoshida:2017swb}.
A key difference between the T-duality of string gas cosmology and the scale factor duality of pre-Big Bang cosmology is that the former requires space to be initially compact, while the latter does not need compactification as the Universe can be infinitely large.

The approach of this paper is to consider a generic universe before the Big Bang, so generally a contracting
universe in the Einstein frame. The goal is to describe the state of matter and the corresponding cosmological
evolution at very high densities, when the energy scale is of the order of the string mass, $M_\mathrm{s}\equiv\ell_\mathrm{s}^{-1}$, from the point of view of string theory.
As the universe contracts, one expects matter that satisfies the usual energy conditions of general relativity
to clump and become inhomogeneous. In fact, the overdensities can be such that matter undergoes collapse
and forms black holes. More precisely, it was shown in Ref.~\cite{Quintin:2016qro} (see also Ref.~\cite{Chen:2016kjx})
with the theory of cosmological perturbations that in a contracting universe hydrodynamical matter with small
sound speed suffers from the Jeans instability and collapses into Hubble-size black holes well before a bounce
is reached. This instability in a generic contracting universe was first studied in Ref.~\cite{Lifshitz:1963},
an analysis that was extended by Ref.~\cite{Banks:2002fe} to argue that the final state of a contracting universe
is a dense gas of black holes with a stiff equation of state (in which the pressure equates the energy density).
In the context of string theory, it was shown in Ref.~\cite{Buonanno:1998bi} that the past-trivial string vacuum of
the tree-level low-energy effective gravidilaton action is also generically prone to gravitational
instability, leading to the formation of black holes. All these studies thus indicate that the state of a
contracting universe at high densities is composed of many black holes.

When the universe reaches the string scale, the black holes are then expected to become more
stringy in nature. In fact, the state of a `black-hole gas' is argued in Ref.~\cite{Veneziano:2003sz} to become a
`string-hole gas'. String holes represent marginal black holes with mass equal to $M_\mathrm{s}g_\mathrm{s}^{-2}$
(see Refs.~\cite{Veneziano:1986zf,Susskind:1993ws} as well as \cite{Veneziano:2003sz,Damour:1999aw,Veneziano:2012yj}), where $g_\mathrm{s}$ is the string coupling.
This represents a correspondence curve along which the physical properties of black holes
and strings match spectacularly well (see, e.g., Refs.~\cite{Damour:1999aw,StringBHcorr}).
In particular, the Schwarzschild radius and Hawking temperature of a string hole are given by the string length and mass, respectively.
Therefore, string holes naturally describe the state of collapsed matter at the string scale. Correspondingly,
a string-hole gas is the logical outcome of a contracting universe in the Einstein frame at high curvature.
The challenge that is tackled in this paper is to find a string-motivated action that can describe the dynamics of
a string-hole gas in agreement with its properties.
In the string frame, Ref.~\cite{Veneziano:2003sz} argued that a string-hole gas should
have vanishing pressure and be described by a constant Hubble parameter and constant dilaton velocity,
though it was not shown explicitly how these properties can arise from a string theory action.

The outline of this paper is as follows.
We first review in Sec.~\ref{sec:SHs} the concept of string holes and carefully derive in Sec.~\ref{sec:SHgas} the properties of a string-hole gas,
both in the Einstein frame and string frame.
We then show in Sec.~\ref{sec:treeleveldilatongrav} that with tree-level dilaton gravity as a low-energy effective action of string theory
dynamics that matches the properties of a string-hole gas is only obtained in finely tuned situations.
It is only when $\alpha'$ corrections are included that we find more appropriate solutions.
We study two different first-order $\alpha'$-corrected actions. First, we extend the work of Ref.~\cite{Gasperini:1996fu}
in Sec.~\ref{sec:alphaprimeGMV} to include the contribution from matter in the dynamical equations.
Second, in Sec.~\ref{sec:alphaprimeMeissner}, we study the $O(d,d)$-invariant action of Ref.~\cite{Meissner:1996sa}.
In Sec.~\ref{sec:phasespace}, we perform a phase space analysis to judge the stability of the string-hole gas solutions
for both $\alpha'$-corrected actions, and we comment on the overall evolutionary scheme. In particular, we address the issue of
connectivity to the string perturbative vacuum.
We summarize the main conclusions in Sec.~\ref{sec:conclusions}. The section is also devoted to a discussion about the possible subsequent fate of a string-hole gas and its role in leading to a nonsingular bouncing cosmology, and we mention future research directions.

Throughout this paper, we work with $\hbar=c=k_\mathrm{B}=1$, and the reduced Planck mass and length are
defined, respectively, by $M_\mathrm{Pl}^{2-D}\equiv 8\pi G$ and $\ell_\mathrm{Pl}\equiv M_\mathrm{Pl}^{-1}$,
where $G$ (also denoted $G_D$) is Newton's gravitational constant in $D=d+1$ spacetime dimensions.
The number of spatial dimensions is denoted by $d$, and we assume that it is an integer greater than or equal to $3$ throughout.

\section{String holes}

\subsection{Black hole/string correspondence}\label{sec:SHs}

One defines a string hole (SH) as an object that has the mass of a Schwarzschild black hole (BH)
confined within a radius given by the string length, i.e.~$M_\mathrm{SH}=M_\mathrm{BH}\sim R_\mathrm{BH}^{D-3}/G$
and $R_\mathrm{SH}=R_\mathrm{BH}=\ell_\mathrm{s}$, so $M_\mathrm{SH}\sim\ell_\mathrm{s}^{D-3}/G$.
(For a review of $D$-dimensional black holes, see, e.g., Ref.~\cite{Emparan:2008eg}).
Introducing the string mass given by the inverse of the string length, $M_\mathrm{s}=\ell_\mathrm{s}^{-1}$,
the string coupling $g_\mathrm{s}$, and the dilaton $\phi$,
we recall the following relation that holds in the weak-coupling regime of the closed string sector (see, e.g., Ref.~\cite{Gasperini:2007zz}):
\begin{equation}
\label{eq:constrantsrelation}
 \left(\frac{\ell_\mathrm{Pl}}{\ell_\mathrm{s}}\right)^{D-2}=\left(\frac{M_\mathrm{s}}{M_\mathrm{Pl}}\right)^{D-2}=g_\mathrm{s}^2=e^\phi\ll 1~.
\end{equation}
From this relation, one can say that a string hole lies along the correspondence curve
\cite{Veneziano:2003sz,Veneziano:1986zf,Susskind:1993ws,Damour:1999aw,Veneziano:2012yj}
\begin{equation}
\label{eq:MSHcorrespondancecurve}
 M_\mathrm{SH}\sim M_\mathrm{s}g_\mathrm{s}^{-2}~.
\end{equation}
It follows that the properties of strings and black holes match impressively
well along this correspondence curve \cite{Damour:1999aw,StringBHcorr}.
For instance, the black hole's Bekenstein-Hawking temperature,
\begin{equation}
 T_\mathrm{BH}=\frac{D-3}{4\pi R_\mathrm{BH}}~,
\end{equation}
and the string's Hagedorn temperature (see, e.g., Ref.~\cite{Atick:1988si}
or \cite{Zwiebach:2004tj} for an introduction),
\begin{equation}
\label{eq:THagdef}
 T_\mathrm{Hag}=\frac{1}{4\pi\sqrt{\alpha'}}~,
\end{equation}
both scale as $\ell_\mathrm{s}^{-1}$ for string holes, where $2\pi\alpha'=\ell_\mathrm{s}^2$.
Similarly, the black hole's Bekenstein-Hawking entropy for a string hole,
\begin{equation}
 S_\mathrm{BH}=\frac{\Omega_{D-2}R_\mathrm{BH}^{D-2}}{4G}\sim\frac{\ell_\mathrm{s}^{D-2}}{\ell_\mathrm{Pl}^{D-2}}\sim g_\mathrm{s}^{-2}~,
\end{equation}
where $\Omega_{D-2}$ is the area of a unit $(D-2)$-sphere,
is of the same order as the entropy of a string,
\begin{equation}
 S_\mathrm{str}=4\pi\sqrt{\alpha'}E\sim\ell_\mathrm{s}M_\mathrm{SH}\sim g_\mathrm{s}^{-2}~,
\end{equation}
where we make use of Eq.~\eqref{eq:MSHcorrespondancecurve} in the last proportionality for a string hole.

From the above correspondence, it is natural to expect a black hole that reaches the size of a fundamental string
to become a string hole. Furthermore, if a contracting universe is populated with a dense gas of black holes,
then the appropriate description of the gas at the string scale must be a string-hole gas.
Hence, the main subject of this paper is the study of a string-hole gas as the state of matter at the string scale
at the end of an Einstein-frame contracting cosmology. The main thermodynamic properties of a string-hole gas are derived in the next subsection.

\subsection{String-hole gas}\label{sec:SHgas}

Let us consider a gas composed of $N$ string holes.
Considering a dense gas, the string holes have negligible momentum,
and the energy of one string hole can be expressed as
$E_\mathrm{SH}=M_\mathrm{SH}\sim \ell_\mathrm{s}^{-1}g_\mathrm{s}^{-2}=\ell_\mathrm{s}^{-1}e^{-\phi}$
by use of Eqs.~\eqref{eq:constrantsrelation} and \eqref{eq:MSHcorrespondancecurve}.
The gas with $N$ string holes thus has total energy
\begin{equation}
 E_\mathrm{gas}\equiv E=NE_\mathrm{SH}\sim N\ell_\mathrm{s}^{-1}e^{-\phi}~.
\end{equation}
In the same way, the entropy of one string hole is $S_{\mathrm{SH}}\sim g_\mathrm{s}^{-2}=e^{-\phi}$, so for a gas of $N$ string holes, one finds
\begin{equation}
 S_\mathrm{gas}\equiv S=NS_{\mathrm{SH}}\sim Ne^{-\phi}~.
\end{equation}

Let the physical volume of the gas be given by $V_\mathrm{gas}\equiv V=fNV_{\mathrm{SH}}$,
where one string hole has volume $V_{\mathrm{SH}}\sim\ell_\mathrm{s}^{D-1}$
and where $f$ is a function that quantifies the separation of the string holes (e.g., $f=1$ for a densely packed string-hole gas,
while $f\gg 1$ for a dilute gas).
Here, we consider a dense gas, so we take $f$ to be of order unity and nearly constant.
Thus, $N\sim V\ell_\mathrm{s}^{1-D}$, and the energy and entropy of the string-hole gas are, respectively, given by
\begin{equation}
\label{eq:ESHgas}
 E\sim V\ell_\mathrm{s}^{-D}e^{-\phi}\sim V\ell_\mathrm{s}^{-2}G^{-1}
\end{equation}
and
\begin{equation}
\label{eq:SSHgas}
 S\sim V\ell_\mathrm{s}^{1-D}e^{-\phi}\sim V\ell_\mathrm{s}^{-1}G^{-1}~,
\end{equation}
where one uses Eq.~\eqref{eq:constrantsrelation} to express $e^\phi\sim G\ell_\mathrm{s}^{2-D}$.
Accordingly, the energy and entropy densities are given by
\begin{align}
\label{eq:rhoSHgas}
 \rho&\equiv\frac{E}{V}\sim\ell_\mathrm{s}^{-D}e^{-\phi}\sim\ell_\mathrm{s}^{-2}G^{-1}~,\\
\label{eq:sdensitySHgas}
 s&\equiv\frac{S}{V}\sim\ell_\mathrm{s}^{1-D}e^{-\phi}\sim\ell_\mathrm{s}^{-1}G^{-1}~,
\end{align}
respectively.

\subsubsection{Einstein-frame properties}

At this point, there are several ways in which one can relate the energy and entropy together.
Let us consider the Einstein frame in which the fundamental constant is Newton's constant,
i.e., $G=\mathrm{constant}$, while the string length can vary as a function of time.
From this point of view, one can eliminate $\ell_\mathrm{s}$ from Eqs.~\eqref{eq:ESHgas} and \eqref{eq:SSHgas}
and relate the energy and entropy through the expression
\begin{equation}
\label{eq:entropyBHgas}
 S\sim\sqrt{\frac{EV}{G}}~,
\end{equation}
or equivalently, from Eqs.~\eqref{eq:rhoSHgas} and \eqref{eq:sdensitySHgas}, the densities are related by $s\sim\sqrt{\rho/G}$.
We note that these equations correspond to the entropy and entropy density equations of a black-hole
gas (see Refs.~\cite{Masoumi:2014vpa,Masoumi:2014nfa,Masoumi:2015sga}
as well as \cite{Banks:2002fe,BanksFischlerHST}).
This makes sense; when viewed in the Einstein frame, the string-hole gas is dominated by its gravitational nature,
i.e., the strings behave more like black holes, at least thermodynamically.

We note that the entropy equation \eqref{eq:entropyBHgas} has been shown \cite{Masoumi:2014vpa}
to be the only formula that is manifestly invariant under the S- and T-dualities
of string theory, at the same time as approaching the standard Bekenstein-Hawking
black-hole entropy at small densities. This entropy expression also appears in different high-energy physics contexts (see Refs.~\cite{Masoumi:2014vpa,Masoumi:2014nfa,Masoumi:2015sga} and references therein).

Using the thermodynamic identity $T^{-1}=(\partial S/\partial E)_V$, keeping $G$ constant
since we are in the Einstein frame, and using Eq.~\eqref{eq:rhoSHgas}, one finds
\begin{equation}
 T\sim\sqrt{\frac{EG}{V}}=\sqrt{\rho G}\sim\ell_\mathrm{s}^{-1}~,
\label{eq:TBHgas}
\end{equation}
and one notes that the temperature is proportional to the Hagedorn temperature \eqref{eq:THagdef}.
Furthermore, using the identity $p=T(\partial S/\partial V)_E$ for the pressure, and using Eq.~\eqref{eq:TBHgas} for the temperature, one finds the equation of state (EOS)
\begin{equation}
 p=\rho~.
\end{equation}
This matches the EOS of a black-hole gas (see Refs.~\cite{Masoumi:2014vpa,Masoumi:2014nfa,Masoumi:2015sga}
as well as \cite{Banks:2002fe,BanksFischlerHST}).

Similarly, if one considers a Friedmann-Lema\^itre-Robertson-Walker (FLRW) universe with scale factor $a$
and if one requires the entropy in a comoving volume $Va^{-d}$ to be constant,
then it follows from Eq.~\eqref{eq:entropyBHgas} that $E\sim V^{-1}\sim a^{-d}$ and furthermore
\begin{equation}
\label{eq:rhoevoEFSHgas}
 \rho\sim a^{-2d}~.
\end{equation}
Consequently, from Eq.~\eqref{eq:rhoSHgas}, this implies
\begin{equation}
\label{eq:aevoEFSHgas}
 a\sim\ell_\mathrm{s}^{1/d}\sim e^{-\frac{\phi}{d(d-1)}}~,
\end{equation}
where one uses again the fact that $G$ is a constant in the Einstein frame.
Therefore, if one considers a string-hole gas in a contracting universe,
then the scale factor, the string length, and the size of the string holes become smaller as time
progresses, while the string coupling, the dilaton, the energy density, and the (Hagedorn) temperature grow.

\subsubsection{String-frame properties}\label{sec:SHgasSFprop}

Let us consider an alternative point of view: the string frame in which the fundamental constant
is the string length, i.e., $\ell_\mathrm{s}=\mathrm{constant}$,
while the gravitational constant can vary as a function of time.
From this point of view, one can eliminate $G$ from Eqs.~\eqref{eq:ESHgas} and \eqref{eq:SSHgas}
and relate the energy and entropy through the expression
\begin{equation}
\label{eq:entropyenergystringgas}
 S\sim\ell_\mathrm{s}E~,
\end{equation}
and equivalently, it follows that $s\sim\ell_\mathrm{s}\rho$.
From $T^{-1}=(\partial S/\partial E)_V$ and keeping the string length constant, it is straightforward to see that
\begin{equation}
 T\sim\ell_\mathrm{s}^{-1}\sim T_\mathrm{Hag}~,
\end{equation}
which is a constant temperature.
Furthermore, from $p=T(\partial S/\partial V)_E$, it follows that
\begin{equation}
 p=0~.
\end{equation}
This confirms the result of Ref.~\cite{Veneziano:2003sz} and again matches what one could have guessed:
in the string frame, the string-hole gas is dominated
by its stringy nature, and this is why the EOS is that of a string gas with equal contribution
from momentum and winding modes (see, e.g., Ref.~\cite{Gasperini:2007zz}).
Also, the expression \eqref{eq:entropyenergystringgas} matches the leading-order behavior of the entropy of a string gas
(see, e.g., Refs.~\cite{Brandenberger:1988aj,Brandenberger:2008nx}).

Similarly, if one requires adiabaticity ($S=\mathrm{constant}$) in a constant comoving volume in FLRW, then it follows that
the energy must be constant; hence,
\begin{equation}
\label{eq:rhoSHgasSF}
 \rho\sim a^{-d}~.
\end{equation}
From the standard conservation equation (more on this in the next section), this is in agreement with an EOS $p=0$.
With Eq.~\eqref{eq:rhoSHgas}, this implies
\begin{equation}
 a\sim G^{1/d}\sim e^{\phi/d}~,
\label{eq:aevoSFSHgas}
\end{equation}
where one uses again the fact that $\ell_\mathrm{s}$ is a constant in the string frame.
Taking the time derivative of the above, this further implies
\begin{equation}
\label{eq:constraintSHgasSF}
 H=\frac{\dot\phi}{d}~,
\end{equation}
where $H\equiv\dot a/a$ is the Hubble parameter and a dot denotes a derivative with respect to the (string-frame) cosmic time $t$.

To be consistent with the fact that the size of the string holes is constant
in the string frame ($R_\mathrm{SH}=\ell_\mathrm{s}=\mathrm{constant}$), there are two possible cosmological evolutionary paths consistent with the constraint \eqref{eq:constraintSHgasSF}.
First, it could be that the universe is static in the string frame ($H=0$), similar to the (quasi)static Hagedorn phase of
string gas cosmology \cite{Brandenberger:1988aj} (see also Refs.~\cite{MoreStringGas,Brandenberger:2011et} for
reviews that highlight the challenges in that context).
Second, it could be that the radius of the string holes is of the order of the Hubble radius ($R_\mathrm{SH}\sim H^{-1}$)
with the string-frame Hubble parameter being constant ($H\sim\ell_\mathrm{s}^{-1}$). In that case, a dense string-hole gas
coincides with having one string hole per Hubble volume.
This last avenue was conjectured in Ref.~\cite{Veneziano:2003sz}
to correspond to the string phase in pre-Big Bang cosmology, and this is what we explore in the rest of this paper.
We note that a dilute gas could also be possible with less than one string hole per Hubble volume in average, but naively, in this situation, curvature would continue to grow until the gas becomes dense. Conversely, an `overdense' gas with more than one string hole per Hubble volume is most likely forbidden by entropy considerations. Indeed, a string-hole gas as defined above exactly saturates the appropriate entropy bound \cite{HoloBound1,HoloBound15,Brustein:1999md,Brustein:2007hd,HoloBound3} (see Refs.~\cite{Veneziano:2003sz,Gasperini:2002bn} and additional references therein). This is also confirmed in the Einstein frame in which saturation occurs when the EOS is $p=\rho$ \cite{Masoumi:2014vpa,BanksFischlerHST}, and this is the only safe outcome with respect to entropy bounds in a contracting FLRW cosmology (see, e.g., Refs.~\cite{Brustein:1999md,Veneziano:2003sz} but also \cite{Masoumi:2014nfa}).
These entropic considerations also reinforce a string-hole gas to be the state of matter at high densities.

In summary, assuming expansion in the string frame, the evolution of a string-hole gas corresponds to a constant Hubble parameter
equal to the string mass, while the dilaton grows linearly with string-frame time
according to the constraint \eqref{eq:constraintSHgasSF}.
We note that expansion in the string frame is consistent with contraction in the Einstein frame; this is shown explicitly in Appendix \ref{sec:EFsol}.
The goal is then to find a string-theoretic effective action that can support the evolution of a string-hole gas,
i.e., an action of which the equations of motion (EOM) have a phase of string-hole gas evolution as a solution.

\section{Dynamics from dilaton gravity}

\subsection{Tree-level dilaton gravity}\label{sec:treeleveldilatongrav}

We first study the string-frame, tree-level, low-energy effective string theory action
(see, e.g., Refs.~\cite{Gasperini:2007zz,Gasperini:2002bn})
\begin{align}
 S_0=&-\frac{1}{2\ell_\mathrm{s}^{d-1}}\int\mathrm{d}^{d+1}x\,\sqrt{|g|}e^{-\phi}\Big(R+g^{\mu\nu}\nabla_\mu\phi\nabla_\nu\phi \nonumber \\
 &+2\ell_\mathrm{s}^{d-1}U(\phi)\Big)~,
\label{eq:S0}
\end{align}
where $g\equiv\mathrm{det}(g^\mu{}_{\nu})$ is the determinant of the metric tensor,
$U(\phi)$ is the potential energy of the dilaton field,
and $R$ denotes the Ricci scalar in this section.
Since we focus on the gravidilaton sector of the effective string theory action,
we set to zero the potential contribution from the antisymmetric field strength
coming from the Neveu-Schwarz/Neveu-Schwarz 2-form.

The above action represents the effective action for vacuum string theory, but we want to consider the addition of matter;
hence, we take the total action to be $S=S_0+S_\mathrm{m}$, where $S_\mathrm{m}$ represents the matter action.
The energy-momentum tensor associated with $S_\mathrm{m}$ is defined as usual by
$T_{\mu\nu}\equiv 2|g|^{-1/2}\delta S_\mathrm{m}/\delta g^{\mu\nu}$. The matter action may also depend on the dilaton,
so
\begin{equation}
 \sigma\equiv -\frac{2}{\sqrt{|g|}}\frac{\delta S_\mathrm{m}}{\delta\phi}
\end{equation}
defines the dilaton (scalar) charge density.

Varying the action \eqref{eq:S0} in a homogeneous, isotropic, and flat FLRW spacetime,
\begin{equation}
 g_{\mu\nu}\mathrm{d}x^\mu\mathrm{d}x^\nu=\mathrm{d}t^2-a(t)^2\delta_{ij}\mathrm{d}x^i\mathrm{d}x^j~,
\end{equation}
a set of dilaton-gravity background EOM
in the string frame can be written as (see, e.g., Refs.~\cite{Gasperini:2007zz,Gasperini:2002bn})
\begin{widetext}
\begin{align}
\label{eq:FriedgenSF}
 d(d-1)H^2+\dot\phi^2-2dH\dot\phi&=2\ell_\mathrm{s}^{d-1}\left(e^\phi\rho+U(\phi)\right)~,\\
\label{eq:Fried2genSF}
 \dot H-H\dot\phi+dH^2&=\ell_\mathrm{s}^{d-1}\left(e^\phi\left(p-\frac{\sigma}{2}\right)-U_{,\phi}\right)~, \\
\label{eq:dilatoneqgenSF}
 2\ddot\phi-\dot\phi^2+2dH\dot\phi-2d\dot H-d(d+1)H^2&=2\ell_\mathrm{s}^{d-1}\left(e^\phi\frac{\sigma}{2}-U(\phi)+U_{,\phi}\right)~,
\end{align}
\end{widetext}
where one assumes that the energy-momentum tensor can be decomposed as a perfect fluid\footnote{We comment on
the possible presence of viscosity as a deviation from a perfect fluid description later in this section.}
with $T^\mu{}_\nu=\mathrm{diag}(\rho,-p\delta^i{}_j)$.
Combining Eqs.~\eqref{eq:FriedgenSF}--\eqref{eq:dilatoneqgenSF}, one can derive the fluid's conservation equation, which goes as
\begin{equation}
\label{eq:conteqgenSF}
 \dot\rho+dH(\rho+p)=\frac{1}{2}\sigma\dot\phi~.
\end{equation}
General power-law solutions to these equations are well known (see, e.g.,
Refs.~\cite{Gasperini:1992em,Gasperini:2007zz,Gasperini:2002bn,Gasperini:2007ar})
but mostly for vanishing potential, vanishing dilaton charge, and an EOS of the form $p=w\rho$.
We want to consider a string-hole gas, in which these assumptions may not all be met.
From Eqs.~\eqref{eq:rhoSHgas} and \eqref{eq:rhoSHgasSF},
a string-hole gas in the string frame has energy density
\begin{equation}
\label{eq:rhoSHgasSFfull}
 \rho=C\ell_\mathrm{s}^{-d-1}e^{-\phi}=\rho_0a^{-d}~,
\end{equation}
where $C$ is a dimensionless positive constant and $\rho_0$ is a positive constant with dimensions of energy density.
As seen in the previous section, this implies the constraint equation $H=\dot\phi/d$. Substituting this constraint and Eq.~\eqref{eq:rhoSHgasSFfull} into the conservation equation \eqref{eq:conteqgenSF}, one finds
\begin{equation}
\label{eq:sigma2p}
 \sigma=2p~,
\end{equation}
independent of the EOS (only assuming $H\neq 0$).
Therefore, one notices that if the dilaton charge density vanishes, the pressure is zero,
which is the naive EOS for a string-hole gas in the string frame as shown in the previous section.
Conversely, if we expect the pressure to vanish from thermodynamic arguments, then this tells us that the string-hole gas matter action
should have no explicit $\phi$ dependence, so the dilaton charge density vanishes.

Inserting the constraint $H=\dot\phi/d=\mathrm{constant}$ (which implies $\ddot\phi=\dot H=0$) and Eq.~\eqref{eq:sigma2p} into Eq.~\eqref{eq:Fried2genSF} immediately yields $U_{,\phi}=0$. Therefore, a fixed-point solution satisfying the constraint $H=\dot\phi/d=\mathrm{constant}$ is only possible with a constant potential independent of the dilaton. Then, Eqs.~\eqref{eq:FriedgenSF} and \eqref{eq:dilatoneqgenSF} further reduce to
\begin{align}
\label{eq:Friedreduced}
 -\frac{d}{2}H^2&=C\ell_\mathrm{s}^{-2}+\ell_\mathrm{s}^{d-1}U~, \\
\label{eq:dilatonreduced}
 -\frac{d}{2}H^2&=wC\ell_\mathrm{s}^{-2}-\ell_\mathrm{s}^{d-1}U~,
\end{align}
where we set the EOS to be of the form $p=w\rho$.
For the above equations to yield a real solution for $H$, the only possibility is to have a constant negative potential,
\begin{equation}
\label{eq:constantUsol}
 U=-\frac{1}{\ell_\mathrm{s}^{d-1}}\left(\frac{d}{2}H_\star^2+\frac{C}{\ell_\mathrm{s}^2}\right)~,
\end{equation}
where the positive constant $H_\star$ should be of the order of $\ell_\mathrm{s}^{-1}$ to yield the solution $H=H_\star\sim\ell_\mathrm{s}^{-1}$.
This is equivalent to introducing a fine-tuned negative cosmological constant, $\Lambda\sim -\mathcal{O}(\ell_\mathrm{s}^{-D})$,
in the string frame\footnote{We note, however, that such a negative constant value of $U$ may naturally appear in the tree-level string effective action, but this would require a noncritical number of dimensions (see, e.g., Ref.~\cite{Gasperini:2007zz}).}.
Any other forms of the potential $U(\phi)$ generically cannot support a string-hole gas evolution with $H=\dot\phi/d=\mathrm{constant}$.
Furthermore, the potential \eqref{eq:constantUsol}, which yields the solution $H_\star$, is only consistent with Eqs.~\eqref{eq:Friedreduced}--\eqref{eq:dilatonreduced} provided the EOS is also tuned to be
\begin{equation}
 w=-1-\frac{d\ell_\mathrm{s}^2H_\star^2}{C}~,
\end{equation}
which violates the null energy condition.
In summary, this avenue does not seem particularly appealing, considering it would require tuning an \textit{ad hoc} negative cosmological constant and the EOS to a physically unexpected value.

This conclusion generalizes to nonlocal potentials of the form $U(\bar\phi)$, where
\begin{equation}
 \bar\phi=\phi-\ln a^d
\label{eq:shifteddilaton}
\end{equation}
is the shifted dilaton.
Indeed, we note that $\dot{\bar\phi}=\dot\phi-dH=0$ for a string-hole gas satisfying the constraint $H=\dot\phi/d$.
Thus, regardless of the modifications to the EOM for a nonlocal potential (see, e.g., Refs.~\cite{Gasperini:2007zz,Gasperini:2002bn,Gasperini:2007ar}
for the exact modified EOM), $\bar\phi$ has to remain constant during a string-hole gas evolution, so any potential
$U(\bar\phi)$ would simply be a constant, i.e., a cosmological constant.

In summary, it appears that one cannot support the evolution of a string-hole gas with tree-level dilaton gravity, no matter the form of the potential
(unless it is a fine-tuned negative cosmological constant). Therefore, one should explore the possibility of higher-order corrections.

\subsection{Action with $\alpha'$ corrections}\label{sec:alphaprimeGMV}

The low-energy effective action $S_0$ introduced in the previous subsection is only compatible with the conformal invariance of
quantized strings on a curved background to zeroth order in $\alpha'\sim\ell_\mathrm{s}^2$. When going to first order,
conformal invariance allows new higher-derivative terms such that the effective action contains terms that scale as
the square of the spacetime curvature and so on. As long as curvature is small, e.g., $\ell_\mathrm{s}^2R\ll1$,
then the perturbative expansion is dominated by the zeroth-order action. However, when the curvature reaches the string scale,
which is the case when $H\sim\ell_\mathrm{s}^{-1}$, then higher-order terms are necessary. In fact, when the perturbative expansion breaks
down on substring scales, working with an effective action is no longer viable, and one would have to work with a proper
conformal field theory that could account for $\alpha'$ corrections nonperturbatively (see, e.g., Ref.~\cite{Kiritsis:1994np}).
This approach, however, is beyond the scope of this study, and in what follows, we assume that a first-order $\alpha'$-corrected effective action
is a sufficient approximation when $H\sim\ell_\mathrm{s}^{-1}$.

Demanding general covariance and gauge invariance of the string effective action, one can write down many perfectly valid
actions that are compatible with the condition of conformal invariance to first order in $\alpha'$.
Those actions are related by simple field redefinitions of the metric and dilaton;
hence, it is ambiguous which action to choose (see, e.g., Refs.~\cite{Gasperini:2002bn,Gasperini:2007zz} and references therein).
For instance, the simplest consistent action to first order in $\alpha'$ is $S=S_0+S_{\alpha'}$ with
\begin{equation}
 S_{\alpha'}=\frac{k\alpha'}{8\ell_\mathrm{s}^{d-1}}\int\mathrm{d}^{d+1}x\,\sqrt{|g|}e^{-\phi}R_{\mu\nu\kappa\lambda}R^{\mu\nu\kappa\lambda}~,
\end{equation}
where $R_{\mu\nu\kappa\lambda}$ is the Riemann tensor
and either $k=1$ for bosonic strings or $k=1/2$ for heterotic superstrings.
However, working with the above action (i.e.~with the square of the Riemann tensor) in a cosmological context is rather cumbersome, because the field equations contain, in general, 
higher than second derivatives of the metric tensor. Such a formal 
complication can be avoided, however, by performing an appropriate field 
redefinition \cite{Gasperini:1996fu} and considering the action with
\begin{equation}
 S_{\alpha'}=\frac{k\alpha'}{8\ell_\mathrm{s}^{d-1}}\int\mathrm{d}^{d+1}x\,\sqrt{|g|}e^{-\phi}\left(\mathcal{G}-(\nabla_\mu\phi\nabla^\mu\phi)^2\right)~,
\label{eq:Salphap}
\end{equation}
where $\mathcal{G}\equiv R_{\mu\nu\kappa\lambda}R^{\mu\nu\kappa\lambda}-4R_{\mu\nu}R^{\mu\nu}+R^2$
is the Gauss-Bonnet invariant, $R_{\mu\nu}\equiv g^{\kappa\lambda}R_{\kappa\mu\lambda\nu}$ is the Ricci
tensor, and $R\equiv g^{\mu\nu}R_{\mu\nu}$ is the Ricci scalar.
This was first considered by Gasperini, Maggiore \& Veneziano \cite{Gasperini:1996fu} (GMV hereafter;
also studied in Refs.~\cite{Brustein:1997cv,Cartier:1999vk,MoreGMV}
and discussed in \cite{Gasperini:2002bn,Gasperini:2007zz}).
Therefore, for a first attempt, we examine the action $S=S_0+S_{\alpha'}+S_\mathrm{m}$ with
$S_{\alpha'}$ given by Eq.~\eqref{eq:Salphap},
and for the rest of this paper, we assume that the dilaton has no potential; i.e., we set $U(\phi)=0$ in $S_0$.

GMV already showed that this action admits no homogeneous and isotropic fixed-point solution with $\dot{\bar\phi}=0$,
i.e.~with $H=\dot\phi/d=\mathrm{constant}$ for a string-hole gas.
However, GMV only considered the vacuum action with no matter, i.e.~$S=S_0+S_{\alpha'}$.
To find dynamics for the string-hole gas, one must include the matter action $S_\mathrm{m}$ as before.
The EOM that follow from varying the corresponding action in a FLRW background are
\begin{widetext}
\begin{align}
 \rho&=\frac{1}{2}\ell_\mathrm{s}^{1-d}e^{-\phi}\left(\dot\phi^2+d(d-1){H}^2
  -2d{H}\dot\phi-\frac{3k\alpha'}{4}\mathcal{F}_{\rho}(H,\dot\phi)\right)~, \nonumber \\
 \sigma&=-\ell_\mathrm{s}^{1-d}e^{-\phi}\left(-2\ddot\phi+2d\dot{H}+\dot\phi^2+d(d+1){H}^2-2d{H}\dot\phi
  +\frac{k\alpha'}{4}\mathcal{F}_{\sigma}(H,\dot\phi,\dot H,\ddot\phi)\right)~, \nonumber \\
 p&=\frac{1}{2d}\ell_\mathrm{s}^{1-d}e^{-\phi}\left(-2d(d-1)\dot{H}+2d\ddot\phi-d^2(d-1){H}^2+2d(d-1){H}\dot\phi-d\dot\phi^2
  +\frac{k\alpha'}{4}\mathcal{F}_{p}(H,\dot\phi,\dot H,\ddot\phi)\right)~,
\label{eq:alphap1EOM3}
\end{align}
where we define
\begin{align}
 \mathcal{F}_{\rho}(H,\dot\phi)\equiv&~c_1{H}^4+c_3{H}^3\dot\phi-\dot\phi^4~, \nonumber \\
 \mathcal{F}_{\sigma}(H,\dot\phi,\dot H,\ddot\phi)\equiv&~3c_3\dot{H}{H}^2-12\ddot\phi\dot\phi^2+(c_1+dc_3){H}^4-4d{H}\dot\phi^3+3\dot\phi^4~, \nonumber \\
 \mathcal{F}_{p}(H,\dot\phi,\dot H,\ddot\phi)\equiv&~12c_1\dot{H}{H}^2+3c_3\ddot\phi{H}^2+6c_3\dot{H}{H}\dot\phi+3dc_1{H}^4
  -2(2c_1-dc_3){H}^3\dot\phi-3c_3{H}^2\dot\phi^2+d\dot\phi^4~,
\end{align}
\end{widetext}
and
\begin{align}
 c_1\equiv&~-\frac{d}{3}(d-1)(d-2)(d-3)~, \nonumber \\
 c_3\equiv&~\frac{4d}{3}(d-1)(d-2)~.
\end{align}
These equations generalize the EOM that were already derived, e.g., in Refs.~\cite{Gasperini:1996fu,Gasperini:2002bn,Gasperini:2007zz},
to include matter; the vacuum limit ($\rho=p=\sigma=0$) reduces to the EOM in Refs.~\cite{Gasperini:1996fu,Gasperini:2002bn,Gasperini:2007zz}.
We note that the above three EOM are not all independent. Indeed, one can verify that the continuity equation
\begin{equation}
\label{eq:continuityeq}
 \dot\rho+d{H}(\rho+p)=\frac{1}{2}\sigma\dot\phi
\end{equation}
relates the three EOM.

We now seek to find solutions to the above EOM that could describe a string-hole gas.
To do so, one sets $\rho=C\ell_\mathrm{s}^{-d-1}e^{-\phi}$, $\sigma=2p$, and $H=\dot\phi/d$.
Furthermore, we relate the pressure and energy density through an EOS of the form $p=w\rho$.
We expect the EOS to be $p=0$ for a string-hole gas in the string frame from the thermodynamic arguments of Sec.~\ref{sec:SHgasSFprop}, so the EOS parameter $w$ is set to zero later on.
Nevertheless, the more crucial property for a string-hole gas is that $p_\mathrm{eff}\equiv p-\sigma/2=0$; thus, we perform a slightly more general analysis in what follows with a generic EOS parameter $w$.
One then looks for fixed-point solutions with $y_1\equiv{H}=\mathrm{constant}$, $y_2\equiv\dot\phi=\mathrm{constant}$,
and $\ddot\phi=\dot{H}=0$. The constraint $H=\dot\phi/d$ implies $y_2=dy_1$,
and the three differential EOM reduce to three algebraic equations for $y_1$,
\begin{align}
 -dy_1^2\left(1-\frac{3k\alpha'y_1^2\Delta}{4}\right)&=2C\ell_\mathrm{s}^{-2}~, \nonumber \\
 dy_1^2\left(1-\frac{k\alpha'y_1^2\Delta}{4}\right)&=-2wC\ell_\mathrm{s}^{-2}~, \nonumber \\
 -d^2y_1^2\left(1-\frac{k\alpha'y_1^2\Delta}{4}\right)&=2dwC\ell_\mathrm{s}^{-2}~,
\end{align}
where we define $\Delta\equiv 2d^2+d-2$.
We note that $\Delta$ is strictly positive (in fact, $\Delta\geq 19$ for $d\geq 3$).
The second and third equations above are completely equivalent,
which is due to the fact that the three EOM are not independent.
Therefore, one only has to solve the first and second equations for $y_1$. Requiring real solutions for $y_1$,
one can show that these two equations yield the same nontrivial solutions,
\begin{equation}
\label{eq:solx1}
 y_1=H=\pm\frac{2}{\ell_\mathrm{s}}\sqrt{\frac{2\pi(1-w)}{k(1-3w)\Delta}}~,
\end{equation}
if and only if $w<1/3$ and
\begin{equation}
\label{eq:rhoamplitudesol1}
 C=\frac{8\pi d(1-w)}{k(1-3w)^2\Delta}~,
\end{equation}
where we use $2\pi\alpha'=\ell_\mathrm{s}^2$ to simplify the expressions.
The solution for $\dot\phi$ immediately follows by multiplying Eq.~\eqref{eq:solx1} by $d$.

A couple of comments are in order.
One first notes that $|H|\sim\ell_\mathrm{s}^{-1}$ as expected.
Second, one notices that the restrictions $w<1/3$ and Eq.~\eqref{eq:rhoamplitudesol1} impose $C>0$,
which means that no real and consistent solution (except the trivial solution $H=\dot\phi=0$) would have followed from setting $C=0$.
This reproduces what was stated by GMV, i.e.~that there exists no consistent nontrivial solution satisfying
the constraint $\dot{\bar\phi}=0$ (which is equivalent to $H=\dot\phi/d=\mathrm{constant}$) in vacuum.
In summary, the GMV $\alpha'$-corrected action that includes a string-hole gas matter action does allow for consistent solutions
with the properties of a string-hole gas for any $w<1/3$ and provided $\rho$ has the appropriate amplitude, with $C$ given in
Eq.~\eqref{eq:rhoamplitudesol1}.

The unique physical solution for the EOS $p=0$ ($w=0$) is then
\begin{equation}
 H=\frac{\dot\phi}{d}=\frac{2}{\ell_\mathrm{s}}\sqrt{\frac{2\pi}{k\Delta}}~,
\label{eq:HsolSFSHgasw0}
\end{equation}
taking the positive solution for expansion in the string frame.
For instance, in $d=3$ dimensions and for $k=1$, the solution is $H=2\ell_\mathrm{s}^{-1}\sqrt{2\pi/19}$.
In the case $w=0$, the physical solution is valid only if $C=8\pi d/(k\Delta)$, which might appear as a fine-tuning problem.
However, we recall that $C$ is only an arbitrary constant amplitude for the energy density [c.f.~Eq.~\eqref{eq:rhoSHgasSFfull}],
and it is certainly tunable depending on the total energy density of the universe
and the other matter contents prior to the string-hole gas phase. In sum,
the $\alpha'$-corrected action considered in this subsection has background EOM that have a unique and natural solution
[Eq.~\eqref{eq:HsolSFSHgasw0}] corresponding to a string-hole gas evolution.

\subsection{$O(d,d)$-invariant $\alpha'$-corrected action}\label{sec:alphaprimeMeissner}

As we mentioned in the previous subsection, there are several consistent $\alpha'$-corrected actions related through
field redefinitions. In this subsection, we consider a different choice for $S_{\alpha'}$, specifically
\begin{align}
 S_{\alpha'}=&~\frac{k\alpha'}{8\ell_\mathrm{s}^{d-1}}\int\mathrm{d}^{d+1}x\,\sqrt{|g|}e^{-\phi}\Big(\mathcal{G}
  -(\nabla_\mu\phi\nabla^\mu\phi)^2 \nonumber \\
  &-4G^{\mu\nu}\nabla_\mu\phi\nabla_\nu\phi+2(\nabla_\mu\phi\nabla^\mu\phi)\Box\phi\Big)~,
\label{eq:Meissneraction}
\end{align}
where $G_{\mu\nu}\equiv R_{\mu\nu}-Rg_{\mu\nu}/2$ is the Einstein tensor
and $\Box\equiv g^{\mu\nu}\nabla_\mu\nabla_\nu$ is the d'Alembertian.
This action shares the Gauss-Bonnet and $(\nabla\phi)^4$ terms with the action \eqref{eq:Salphap},
but the second in line in Eq.~\eqref{eq:Meissneraction} is new; nevertheless, this action is still free from higher derivatives in the cosmological field equations.
The actions \eqref{eq:Salphap} and \eqref{eq:Meissneraction} are related by a field redefinition (see Ref.~\cite{Gasperini:2007zz}).
This action was first introduced by Meissner \cite{Meissner:1996sa} (see also
Ref.~\cite{MoreMeissner})
who showed that it is invariant under the $O(d,d)$ symmetry to first order in the $\alpha'$ expansion.

The $O(d,d)$ symmetry plays a key role in string theory and even more in the context of pre-Big Bang cosmology
(see Refs.~\cite{Gasperini:2002bn,Gasperini:2007zz} and references therein).
Indeed, the cosmological scale factor duality $a\rightarrow 1/a$ \cite{Veneziano:1991ek}
is actually extendable to a continuous symmetry, the transformation group of which is $O(d,d)$. It was found that the action of the group
transforms known solutions to the effective cosmological
string theory into new solutions \cite{MoreOdd,Sen:1991zi}.
The symmetry was shown to be present for the low-energy action to zeroth order
in $\alpha'$ with the presence of matter \cite{Gasperini:1991ak}, but it was also
argued to apply to all orders in $\alpha'$ \cite{Veneziano:1991ek,Sen:1991zi}.
The action that has the symmetry to first order in $\alpha'$ is the one found by Meissner \cite{Meissner:1996sa},
and it is the one introduced above that we consider below.

Since $S_0$ is already invariant under $O(d,d)$ transformations \cite{Gasperini:1991ak,Gasperini:2002bn,Gasperini:2007zz},
it is natural to consider the $\alpha'$-corrected action \eqref{eq:Meissneraction} that bares the same symmetry.
Let us comment on the nature of the symmetry for a string-hole gas.
Considering an isotropic and homogeneous cosmology for simplicity, the EOM of the full action $S=S_0+S_{\alpha'}+S_\mathrm{m}$
are $O(d,d)$ invariant under the transformations $a\rightarrow 1/a$, $\bar\phi\rightarrow\bar\phi$, $\bar\rho\rightarrow\bar\rho$,
$\bar p\rightarrow -\bar p$, and $\bar\sigma\rightarrow\bar\sigma$, where the shifted dilaton is given by Eq.~\eqref{eq:shifteddilaton}
and the other shifted variables are $\bar\rho=\rho a^d$, $\bar p=pa^d$, and $\bar\sigma=\sigma a^d$.
Thus, for a string-hole gas with $p=\sigma/2=0$, we expect $\bar\rho=C\ell_\mathrm{s}^{-d-1}e^{-\bar\phi}=\rho_0$, $\bar p=0$, and $\bar\sigma=0$,
and readily, we notice the $O(d,d)$ invariance.
Let us mention that in general, though, deviations from a perfect fluid description could change this conclusion.
Indeed, it was shown in Ref.~\cite{Gasperini:1991ak} that a particular nontrivial action of the $O(d,d)$ group
can transform a perfect fluid with a diagonal stress tensor into a fluid with nondiagonal elements in its stress tensor.
More precisely, a perfect fluid with EOS $p=w\rho$ transforms into a pressureless fluid
(so $p\rightarrow 0$) with shear viscosity given by $\eta=-w\rho/(2H)$.
However, for a string-hole gas, the perfect fluid EOS is precisely expected to be that of a pressureless fluid to start with ($w=0$),
so the transformation turns out to be trivial, and no shear viscosity appears.
Therefore, a string-hole gas with vanishing pressure in the string frame has a valid and consistent perfect fluid description
from the point of view of $O(d,d)$ invariance of its action.
If one allows $w\neq 0$ to describe a string-hole gas (but still with $p_\mathrm{eff}=p-\sigma/2=0$), then a more refined analysis should drop the perfect fluid description and include the possible effects of viscosity, as was first considered in Ref.~\cite{Masoumi:2014nfa}. We keep the exploration of this possibility for future work.

Let us now derive the EOM. We consider the FLRW metric
\begin{equation}
 g_{\mu\nu}\mathrm{d}x^\mu\mathrm{d}x^\nu=N(t)^2\mathrm{d}t^2-e^{2\beta(t)}\delta_{ij}\mathrm{d}x^i\mathrm{d}x^j~,
\end{equation}
where, in this subsection, we introduce the lapse function $N(t)$ [which we later set to $N(t)\equiv 1$].
Also, the scale factor is written as $a(t)=e^{\beta(t)}$, so the Hubble parameter becomes $H(t)=\dot\beta(t)$.
This is only a matter of convenience to compute the EOM below.
The action $S=S_0+S_{\alpha'}$ thus reduces to the form
$S=-(\ell_\mathrm{s}/2)\int\mathrm{d}t~\mathcal{V}_t\mathcal{L}(t)$,
where
$\mathcal{V}_t\equiv\ell_\mathrm{s}^{-d}\int_{\Sigma_t}\mathrm{d}^dx$
is the volume of the spatial hypersurface of constant time $\Sigma_t$ (at time $t$) in string units, and the Lagrangian density is
\begin{widetext}
\begin{align}
 \mathcal{L}(t)=&~e^{d\beta-\phi}\Big\{\frac{1}{N}\Big[-2d\ddot\beta-d(d+1)\dot\beta^2+2dF\dot\beta+\dot\phi^2\Big] \nonumber \\
 &-\frac{k\alpha'}{4N^3}
 \Big[-3c_3F\dot\beta^3+(d+1)d(d-1)(d-2)\dot\beta^4+3c_3\ddot\beta\dot\beta^2-2d(d-1)\dot\phi^2\dot\beta^2+2\ddot\phi\dot\phi^2
 +2d\dot\phi^3\dot\beta-2F\dot\phi^3-\dot\phi^4\Big]\Big\}~,
\end{align}
where $F\equiv\dot N/N$. After integration by parts, the action reduces to
\begin{equation}
 S=\frac{\ell_\mathrm{s}}{2}\int\mathrm{d}t~\mathcal{V}_te^{d\beta-\phi}\Big\{\frac{1}{N}\Big[-\dot\phi^2-d(d-1)\dot\beta^2+2d\dot\beta\dot\phi\Big]
 +\frac{k\alpha'}{4N^3}\Big[c_1\dot\beta^4+c_3\dot\phi\dot\beta^3-2d(d-1)\dot\phi^2\dot\beta^2+\frac{4}{3}d\dot\phi^3\dot\beta-\frac{1}{3}\dot\phi^4\Big]\Big\}~.
\end{equation}
Let us add to the above action a matter action $S_\mathrm{m}$ described by
an energy density $\rho$, pressure $p$, and dilaton charge density $\sigma$ as before.
Then, varying the total action with respect to $N$, $\phi$, and $\beta$ [and afterward setting $N(t)\equiv 1$], one finds three EOM,
which are the same as the set of equations \eqref{eq:alphap1EOM3},
except the functions $\mathcal{F}_{\rho}$, $\mathcal{F}_{\sigma}$, and $\mathcal{F}_{p}$ that are replaced by
\begin{align}
 \mathcal{F}_{\rho}(H,\dot\phi)=&~c_1{H}^4+c_3{H}^3\dot\phi-2d(d-1){H}^2\dot\phi^2+\frac{4}{3}d{H}\dot\phi^3-\frac{1}{3}\dot\phi^4~, \\
 \mathcal{F}_{\sigma}(H,\dot\phi,\dot H,\ddot\phi)=&~3c_3{\dot H}{H}^2-8d(d-1){\dot H}{H}\dot\phi+4d{\dot H}\dot\phi^2-4d(d-1)\ddot\phi{H}^2
  +8d\ddot\phi{H}\dot\phi-4\ddot\phi\dot\phi^2+(c_1+dc_3){H}^4 \nonumber \\
    &-4d^2(d-1){H}^3\dot\phi+2d(3d-1){H}^2\dot\phi^2-4d{H}\dot\phi^3+\dot\phi^4~, \label{eq:FphiM} \\
 \mathcal{F}_{p}(H,\dot\phi,\dot H,\ddot\phi)=&~12c_1{\dot H}{H}^2+6c_3{\dot H}{H}\dot\phi-4d(d-1){\dot H}\dot\phi^2+3c_3\ddot\phi{H}^2
  -8d(d-1)\ddot\phi{H}\dot\phi+4d\ddot\phi\dot\phi^2+3dc_1{H}^4 \nonumber \\
    &-2(2c_1-dc_3){H}^3\dot\phi-(3c_3+2d^2(d-1)){H}^2\dot\phi^2+4d(d-1){H}\dot\phi^3-d\dot\phi^4~. \label{eq:FbetaM}
\end{align}
\end{widetext}
Note that we reexpressed the Hubble parameter $\dot\beta$ with $H$.

As in the previous subsection, we consider a string-hole gas with
$\rho=C\ell_\mathrm{s}^{-d-1}e^{-\phi}$, $\sigma=2p$, $p=w\rho$, and ${H}=\dot\phi/d$.
One looks for fixed-point solutions with $y_1\equiv{H}=\mathrm{constant}$, $y_2\equiv\dot\phi=\mathrm{constant}$
(so ${\dot H}=\ddot\phi=0$), and $y_2=dy_1$. The three EOM then reduce to two independent algebraic equations:
\begin{align}
 -dy_1^2\left(1-\frac{3(d-2)k\alpha'}{4}y_1^2\right)&=\frac{2C}{\ell_\mathrm{s}^2}~; \\
 dy_1^2\left(1-\frac{(d-2)k\alpha'}{4}y_1^2\right)&=-\frac{2wC}{\ell_\mathrm{s}^2}~.
\end{align}
Those two equations share the same nontrivial solutions,
\begin{equation}
\label{eq:soly1genMeissner}
 y_1={H}=\pm\frac{2}{\ell_\mathrm{s}}\sqrt{\frac{2\pi(1-w)}{k(d-2)(1-3w)}}~,
\end{equation}
if and only if the amplitude parameter satisfies
\begin{equation}
\label{eq:solCgenMeissner}
 C=\frac{8\pi d}{k(d-2)}\frac{1-w}{(1-3w)^2}
\end{equation}
and as long as $w<1/3$.
These expressions are not the same as Eqs.~\eqref{eq:solx1} and \eqref{eq:rhoamplitudesol1},
but they only differ by numerical factors that depend on the number of spatial dimensions. Essentially, $\Delta=2d^2+d-2$ in Eqs.~\eqref{eq:solx1} and \eqref{eq:rhoamplitudesol1} is replaced by $d-2$ in Eqs.~\eqref{eq:soly1genMeissner} and \eqref{eq:solCgenMeissner}.
The solutions are certainly of the same order, and as before (as expected), $|H|\sim\ell_\mathrm{s}^{-1}$.
The physical solution with $w=0$ reduces to
\begin{equation}
 H=\frac{\dot\phi}{d}=\frac{2}{\ell_\mathrm{s}}\sqrt{\frac{2\pi}{k(d-2)}}~,
\label{eq:HsolSFSHgasw02}
\end{equation}
and it requires $C=8\pi d/(k(d-2))$. As before, we argue that $C$ is an arbitrary constant, so this does not represent fine-tuning.
Therefore, the $O(d,d)$-invariant $\alpha'$-corrected action of this subsection yields a unique and natural solution,
which corresponds to a string-hole gas evolution but which is different from the solution of the previous subsection.
The differences are due to the fact that the physical effects of the 
higher-curvature corrections are not invariant, in general, under 
field redefinitions truncated to first order in $\alpha'$. Such an 
ambiguity affects all models truncated to any given finite order of the 
$\alpha'$ expansion and can be resolved, in principle, only by 
considering exact conformal models, which automatically include the 
corrections to all orders.
In the following section, restricting our discussion to the first order 
in $\alpha'$, we perform a phase space analysis of the two previous 
solutions in order to find the most appropriate one to describe --- in 
this approximation --- the main properties of the string-hole gas and of 
its dynamical evolution.

\section{Phase space analysis}\label{sec:phasespace}

At this point, two distinct solutions that correspond to a string-hole gas evolution given two different
$\alpha'$-corrected actions have been found: the solutions \eqref{eq:HsolSFSHgasw0} and \eqref{eq:HsolSFSHgasw02} follow from GMV's action and Meissner's action,
respectively. In the perspective of a greater evolutionary scheme, we now seek to determine the stability of those fixed points
in the whole phase space of cosmological solutions.
For instance, the nontrivial fixed points found by GMV \cite{Gasperini:1996fu} in vacuum were shown to be attractors
in phase space and smoothly connected to the string perturbative vacuum
(i.e., to the asymptotic state with vanishing string coupling and flat 
spacetime, $g_\mathrm{s}\rightarrow 0$ and $H\rightarrow 0$).
Conversely, the attractor fixed points from Meissner's action in vacuum are disconnected from the low-energy trivial fixed point
(see Refs.~\cite{Gasperini:2002bn,Gasperini:2007zz}).
We analyze the phase space with the addition of matter and in particular for a string-hole gas in the subsequent subsections.

\subsection{Stability of the fixed point with GMV's action}\label{sec:stabfixedpointGMV}

Recall the GMV EOM given by Eq.~\eqref{eq:alphap1EOM3}. In general, for an EOS of the form $p=w\rho$ and assuming that $\sigma$ is also proportional to $\rho$, one can see that there are only three independent variables in configuration space: $H$, $\dot\phi$, and $e^\phi\rho$. One can choose to use the Hamiltonian constraint [the first equation of the set \eqref{eq:alphap1EOM3}] to eliminate $e^\phi\rho$ from the other two evolution equations. This amounts to projecting the configuration space onto a two-dimensional vector space, where the vectors are of the form $y^A=(H,\dot\phi)$, $A\in\{1,2\}$. One can thus reexpress the set \eqref{eq:alphap1EOM3} as two independent differential equations, written in vector form as $\dot{y}^A=(\dot H,\ddot\phi)=\mathcal{C}^A$, where $\mathcal{C}^1$ and $\mathcal{C}^2$ are functions of $H$ and $\dot\phi$ only [i.e.~$\mathcal{C}^A=\mathcal{C}^A(y^B)$].
For example, when $p=\sigma/2=0$ ($w=0$), their expressions are
\begin{widetext}
\begin{align}
 \mathcal{C}^1({H},\dot\phi)=&-\frac{1}{\mathcal{D}}\bigg\{16{H}(d{H}-\dot\phi)+\frac{2k\ell_\mathrm{s}^2}{\pi}\Big[\frac{3c_3}{4}(d+3){H}^4-\frac{3c_3}{2d}{H}^3\dot\phi-2(d-1)(2d-1){H}^2\dot\phi^2+2(2d-3){H}\dot\phi^3-\dot\phi^4\Big] \nonumber \\
  &+\frac{3k^2\ell_\mathrm{s}^4}{4\pi^2}\Big[\frac{3c_3^2}{16d}(d+1){H}^6+3c_1{H}^4\dot\phi^2+\frac{c_3}{d}(2d-3){H}^3\dot\phi^3-\frac{9c_3}{4d}{H}^2\dot\phi^4+\dot\phi^6\Big]\bigg\}~, \label{eq:C1}
\end{align}
\begin{align}
 \mathcal{C}^2({H},\dot\phi)=&-\frac{1}{\mathcal{D}}\bigg\{8\Big[d(d-1){H}^2-\dot\phi^2\Big]+\frac{k\ell_\mathrm{s}^2}{\pi}\Big[\frac{3c_3}{4}(d-7){H}^4+\frac{3c_3}{2d}(2d-3){H}^2\dot\phi^2-8(d-1)^2{H}\dot\phi^3+(4d-3)\dot\phi^4\Big] \nonumber \\
  &-\frac{3k^2\ell_\mathrm{s}^4c_3}{16\pi^2d}{H}\Big[3c_1{H}^5+6c_1{H}^4\dot\phi+3c_3{H}^3\dot\phi^2+4d(d-3){H}^2\dot\phi^3-3(4d-3){H}\dot\phi^4+6\dot\phi^5\Big]\bigg\}~, \label{eq:C2}
\end{align}
\end{widetext}
where
\begin{align}
 \mathcal{D}\equiv &~16+\frac{3k\ell_\mathrm{s}^2c_3}{\pi d}\left((d+3){H}^2+2{H}\dot\phi-\frac{3}{d-2}\dot\phi^2\right) \nonumber \\
    &+\frac{3k^2\ell_\mathrm{s}^4c_3}{4\pi^2d}{H}\left(\frac{3c_3}{4}{H}^3-3(d-3){H}\dot\phi^2+6\dot\phi^3\right)~.
\end{align}
For $w=0$, we recall that the string-hole gas fixed point is given by Eq.~\eqref{eq:HsolSFSHgasw0}, and here we denote it as
\begin{equation}
\label{eq:fixedpoint1}
 y^A_\star=(H_\star,\dot\phi_\star)=\frac{2}{\ell_\mathrm{s}}\sqrt{\frac{2\pi}{k\Delta}}(1,d)~.
\end{equation}
One can check that $\mathcal{C}^1({H}_\star,\dot\phi_\star)=\mathcal{C}^2({H}_\star,\dot\phi_\star)=0$ ($\mathcal{C}^A(y^B_\star)=0$), so
$\dot{H}_\star=\ddot\phi_\star=0$ ($\dot{y}^A_\star=0$) as expected.

The Jacobian matrix for the system of differential equations is then
\begin{equation}
\label{eq:Jacobian}
 J_A{}^B=\partial_A\mathcal{C}^B~,
\end{equation}
and its eigenvalues are
\begin{align}
 r_\pm=&~\frac{1}{2}\bigg\{\partial_{{H}}\mathcal{C}^1+\partial_{\dot\phi}\mathcal{C}^2
 \pm\Big[\big(\partial_{{H}}\mathcal{C}^1+\partial_{\dot\phi}\mathcal{C}^2\big)^2 \nonumber \\
 &-4\big(\partial_{{H}}\mathcal{C}^1\partial_{\dot\phi}\mathcal{C}^2
 -\partial_{\dot\phi}\mathcal{C}^1\partial_{{H}}\mathcal{C}^2\big)\Big]^{1/2}\bigg\}~.
\end{align}
After calculating the partial derivatives and evaluating at the fixed point $({H}_\star,\dot\phi_\star)$, one finds
\begin{equation}
 r_\pm=\pm\frac{2}{\ell_\mathrm{s}}\sqrt{\frac{2\pi d\Delta}{k\mathcal{Q}}}~,
\end{equation}
where
\begin{equation}
 \mathcal{Q}\equiv 16d^5-32d^4-46d^3+47d^2+36d-20~.
\end{equation}
Since $r_+>0>r_-$, it follows that the fixed point $({H}_\star,\dot\phi_\star)$ is a saddle point, and therefore, it is generally not stable and certainly not an attractor in phase space.

\begin{figure*}
\centering
 \includegraphics[scale=0.84]{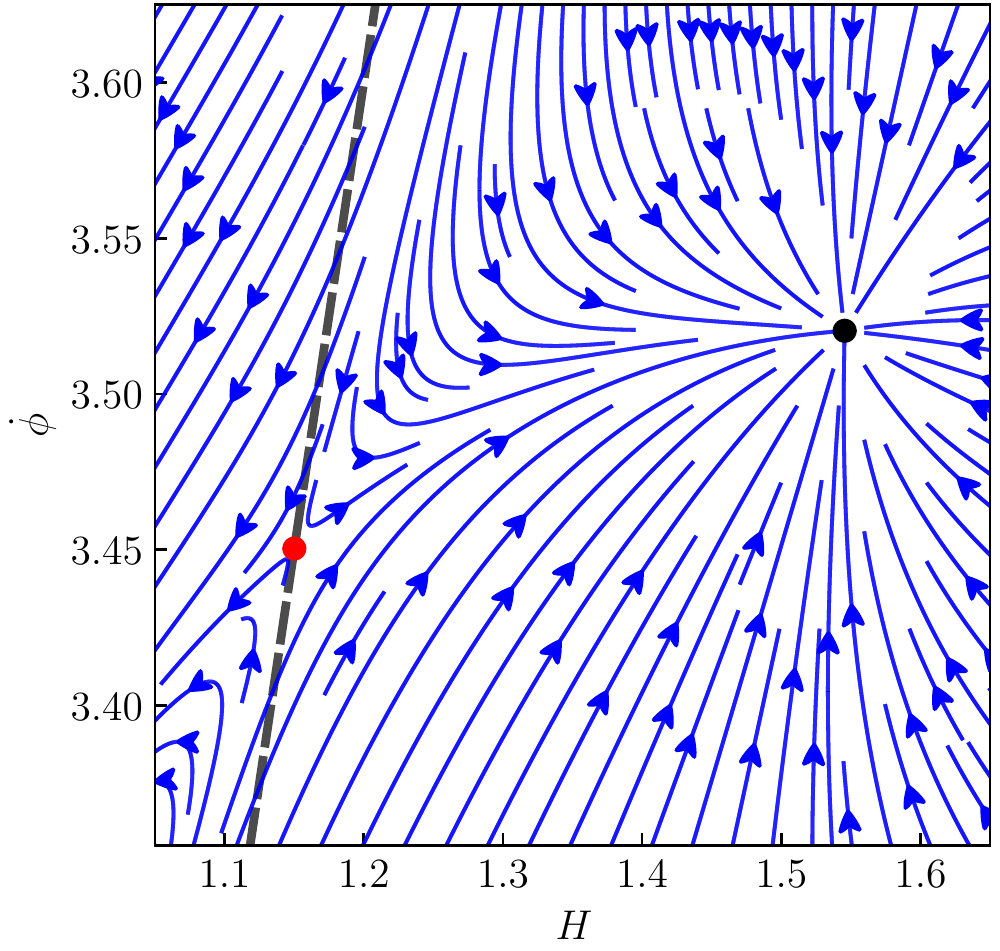}\hspace*{0.5cm}
 \includegraphics[scale=0.84]{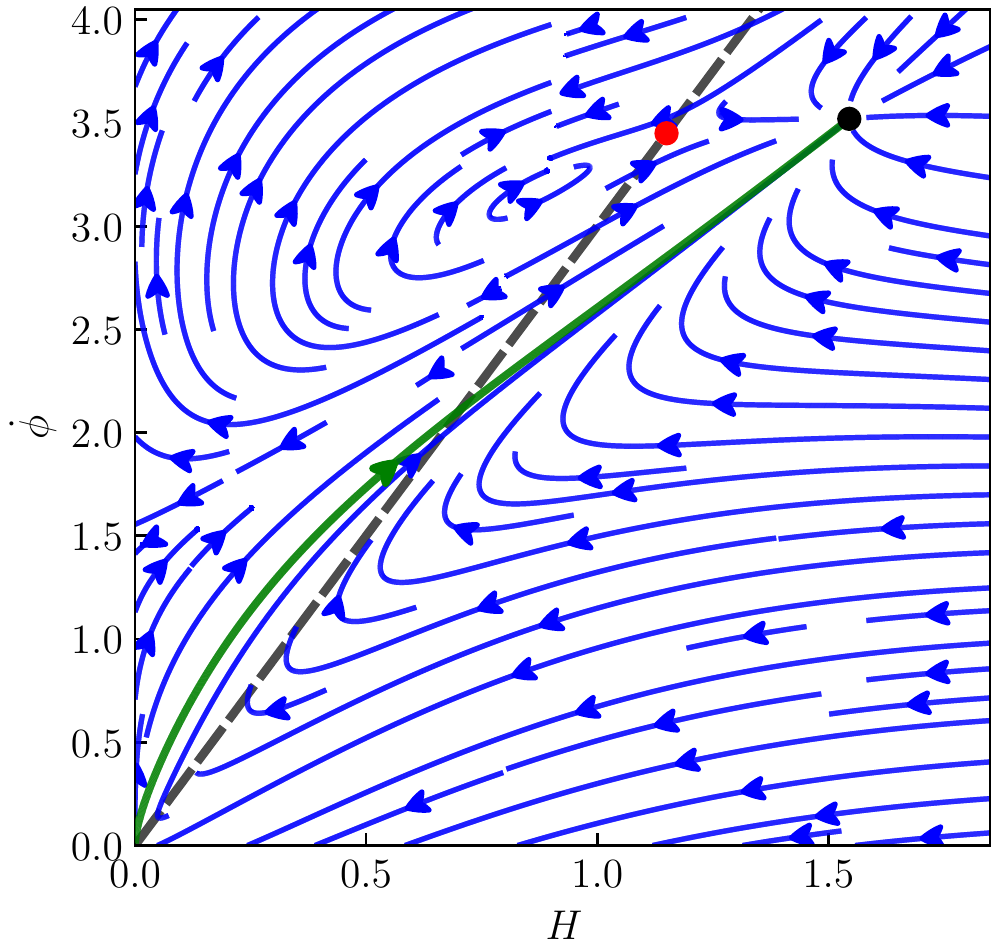}
 \caption{Phase space trajectories for GMV's action in a FLRW background with matter satisfying the continuity equation and $p=\sigma/2=0$.
 Setting $k=1$, $\ell_\mathrm{s}=1$, and $d=3$, $\dot H$ and $\ddot\phi$ are computed from Eqs.~\eqref{eq:C1} and \eqref{eq:C2}, respectively.
 The red dot denotes the string-hole gas saddle point \eqref{eq:fixedpoint1},
 and the black dot denotes the attractor fixed point of vacuum pre-Big Bang cosmology.
 The dashed gray curve depicts the line $\dot\phi=dH$, along which the saddle point is stable.
 The left and right plots show different ranges in ${H}$ and $\dot\phi$.
 The left plot is a blowup of the right plot near the two nontrivial fixed points.
 In the right plot, the green line shows an example of trajectory that starts near the trivial fixed point at ${H}=\dot\phi=0$
 and goes to the attractor fixed point.}
\label{fig:ps1}
\end{figure*}

If one worries only about perturbations around the fixed point that preserve the condition $H=\dot\phi/d$, one may check
the directional derivative of $\mathcal{C}^A$ with respect to the unit vector parallel to the line corresponding to $H=\dot\phi/d$.
The unit vector is expressed as $u^A=(1+d^2)^{-1/2}(1,d)$. The expression for the directional derivative is then
\begin{equation}
 \bm{D}_{\bm{u}}\mathcal{C}^A\equiv u^B\partial_B\mathcal{C}^A
 =\frac{\partial_{{H}}\mathcal{C}^A+d\partial_{\dot\phi}\mathcal{C}^A}{\sqrt{1+d^2}}~,
\end{equation}
and upon calculating the partial derivatives and evaluating at the fixed point $({H}_\star,\dot\phi_\star)$, one finds
\begin{align}
 \left.\bm{D}_{\bm{u}}\mathcal{C}^1\right|_{({H}_\star,\dot\phi_\star)}
  &=-\frac{\mathcal{P}^1}{\ell_\mathrm{s}\mathcal{Q}}\sqrt{\frac{2\pi}{k(2d^4+d^3+d-2)}}~, \nonumber \\
 \left.\bm{D}_{\bm{u}}\mathcal{C}^2\right|_{({H}_\star,\dot\phi_\star)}
  &=-\frac{\mathcal{P}^2}{\ell_\mathrm{s}\mathcal{Q}}\sqrt{\frac{2\pi\Delta}{k(1+d^2)}}~,
\end{align}
where
\begin{align}
 \mathcal{P}^1&\equiv 4d(d+2)(2d-1)\Delta~, \nonumber \\
 \mathcal{P}^2&\equiv 8d^4-8d^3-18d^2+20d~,
\end{align}
Noting that $\mathcal{P}^A>0$ and $\mathcal{Q}>0$ for any $d\geq 3$, it follows that
\begin{equation}
 \left.\bm{D}_{\bm{u}}\mathcal{C}^A\right|_{({H}_\star,\dot\phi_\star)}<0~,\qquad A=1,2~,
\end{equation}
and thus, the fixed point $({H}_\star,\dot\phi_\star)$ is stable in the direction corresponding to the line $H=\dot\phi/d$.
This implies that if one considers perturbations about the string-hole gas saddle point
that respect the condition $H=\dot\phi/d$
the string-hole gas evolution is stable.
However, for general perturbations about the saddle point, the trajectories might flow
away from the string-hole gas evolution.

Further insight can be gained numerically. For example, setting $k=1$, $\ell_\mathrm{s}=1$, and $d=3$, one finds two real
positive nontrivial fixed points that satisfy
$\mathcal{C}^A({H},\dot\phi)=0$: the string-hole gas fixed point with $y_\star^A=(2\sqrt{2\pi/19},6\sqrt{2\pi/19})$
and another fixed point approximately located at $(1.546,3.520)$.
The phase space trajectories are plotted in Fig.~\ref{fig:ps1}. The string-hole gas fixed point is depicted by the red dot,
and visual inspection confirms that it is a saddle point (see the left plot of Fig.~\ref{fig:ps1} for a close-up).
The other fixed point, depicted by the black dot, is the attractor
of standard (vacuum) pre-Big Bang cosmology\footnote{Our numerical values differ from those of
Refs.~\cite{Gasperini:1996fu,Gasperini:2002bn,Gasperini:2007zz} simply due to the choice of units. We work with
$k=\ell_\mathrm{s}=1$, while Refs.~\cite{Gasperini:1996fu,Gasperini:2002bn,Gasperini:2007zz} set $k\alpha'=1$,
so basically the numbers differ by a factor of $\sqrt{2\pi}$.}
(see, e.g., Refs.~\cite{Gasperini:1996fu,Gasperini:2002bn,Gasperini:2007zz}).
We note that this is \emph{exactly} the fixed point found by GMV, and it appears in the phase space no matter what the EOS parameter $w$ is since $e^\phi\rho\rightarrow 0$ at that point.

The dashed gray curves in Fig.~\ref{fig:ps1} depict the line $\dot\phi=dH$. When projecting the trajectories onto that line, it is clear from the left plot that the flow is attracted toward the string-hole gas saddle point in its vicinity. This is in agreement with the earlier (analytical) result that the string-hole gas saddle point is stable in the direction of the constraint $H=\dot\phi/d$.

In the right plot of Fig.~\ref{fig:ps1}, we show the phase space including the trivial fixed point $(H,\dot\phi)=(0,0)$
corresponding to the string perturbative vacuum, and the green curve shows one trajectory passing infinitesimally close to that fixed point.
We notice that it smoothly reaches the attractor fixed point (black dot), confirming the result of GMV\footnote{This time, we note that this curve may not be \emph{exactly} the solution found by GMV. However, it is close enough since $e^\phi\rho$ is subdominant at all times along the green trajectory. In particular, it shares its qualitative behavior: the perturbative evolution starts in the region $\dot{\bar{\phi}}=\dot{\phi}-dH>0$ (above the dashed gray line), crosses the gray line (where $\dot{\bar{\phi}}=0$), and ends at the attractor fixed point in the region $\dot{\bar{\phi}}<0$ (below the gray line).} \cite{Gasperini:1996fu}.
This also implies, however, that it is not possible for a trajectory to start near the string perturbative vacuum and evolve toward
the string-hole gas fixed point smoothly. In the context of pre-Big Bang cosmology, the goal would be to start at the string perturbative vacuum and evolve toward a string-hole gas
as the high-energy state of the universe before a bounce. Although GMV's $\alpha'$-corrected action
allows for a unique string-hole gas solution, it does not seem to be sufficient to describe the evolution of the universe
thoroughly from the perturbative vacuum to the stringy state at high energies. This is not surprising because black-/string-hole formation is not a continuous process; rather, the holes collapse instantaneously from the vacuum fluctuations that have grown in amplitude. Therefore, asking for continuous trajectories connecting the vacuum to the string-hole gas fixed point is ill posed.

Nevertheless, there are arguments to support that a string-hole gas should be connected to the vacuum in some way. In a broader cosmological context, one could imagine starting asymptotically far in the past in a contracting universe (in the Einstein frame)
which has `normal' matter (e.g., a mix of dust [$w=0$] and radiation [$w=1/d$]). As shown in Ref.~\cite{Quintin:2016qro}
(see also Refs.~\cite{Chen:2016kjx,Banks:2002fe,Lifshitz:1963}), starting with vacuum initial conditions,
the pressureless matter would collapse into a black-hole gas, and as stated in the present work, it would evolve into a string-hole gas
with EOS $p=\rho$. From that point of view, a string-hole gas with EOS $w=1$ is naturally an attractor\footnote{Matter with the EOS $w=1$
is generally (marginally) an attractor in a contracting universe, whether it is a black-/string-hole gas,
anisotropies, or a massless scalar field. We use the word `marginal' since any other component with EOS $w>1$,
e.g., an Ekpyrotic field with negative exponential potential, would overturn this conclusion and become the new attractor
(see, e.g., Ref.~\cite{Heard:2002dr}).}, and the same conclusion would necessarily follow in the string frame,
although the physical intuition might be less obvious in the string frame.
In that context, one cannot describe the entire cosmological evolution with the stringy actions studied in this paper; they would be applicable only at the time of formation of the string holes.
In that case, when the condition $H=\dot\phi/d$ is met, as we showed above, the string-hole gas evolution is an attractor in the string frame.

\subsection{Stability of the fixed point with Meissner's action}

\begin{figure*}
\centering
 \includegraphics[scale=0.84]{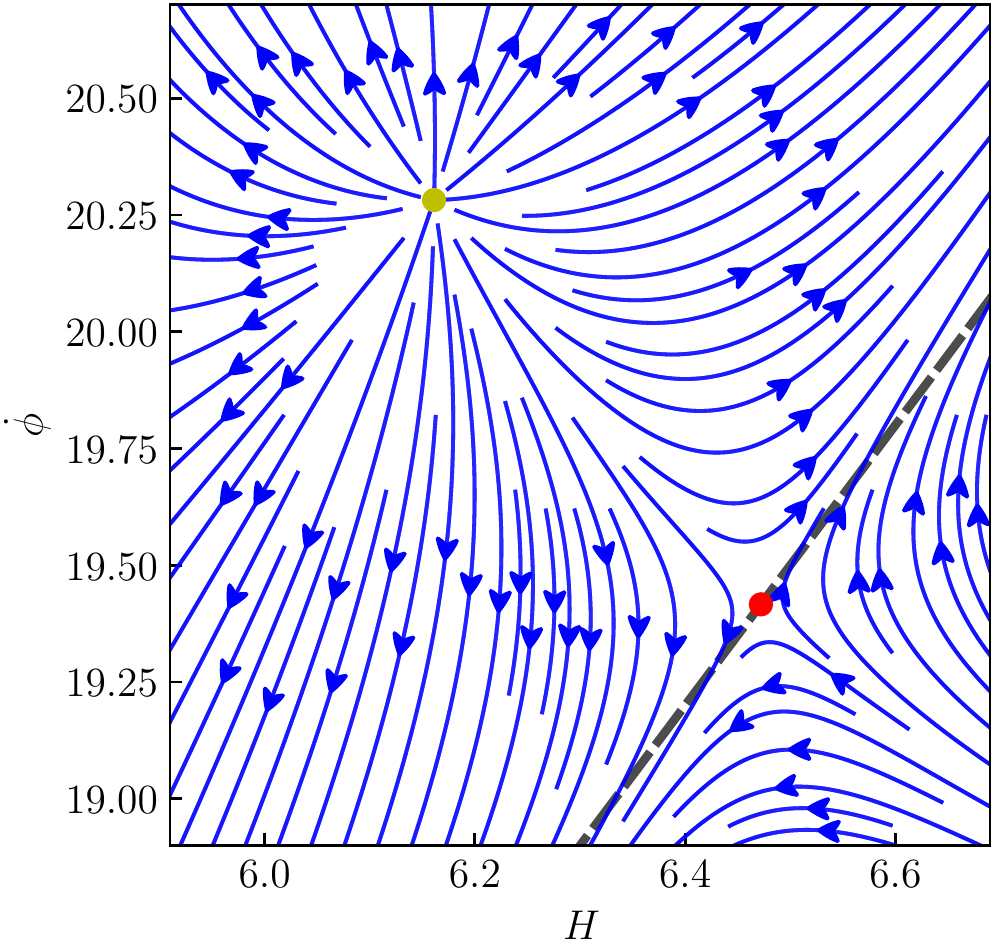}\hspace*{0.5cm}
 \includegraphics[scale=0.84]{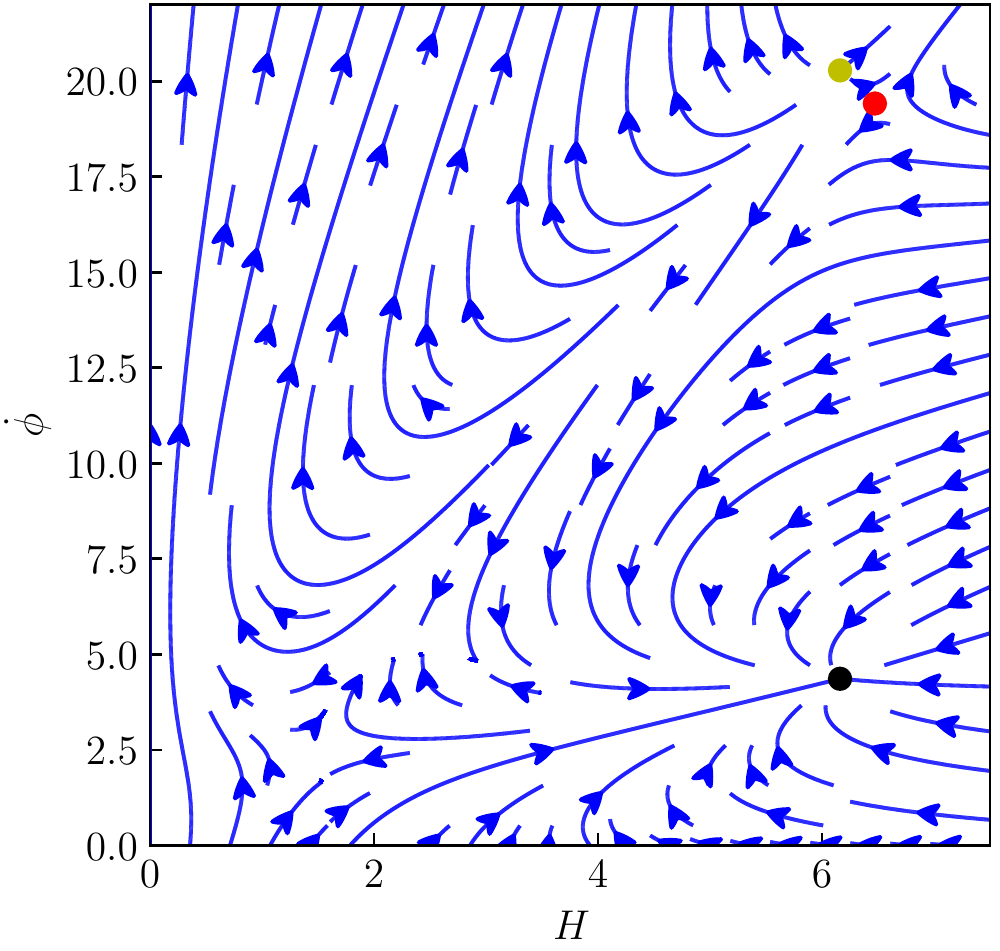}
 \caption{Phase space trajectories for Meissner's action in a FLRW background with matter satisfying the continuity equation and $p=\sigma/2=0$
 (and setting $k=1$, $\ell_\mathrm{s}=1$, and $d=3$).
 The red dot denotes the string-hole gas saddle point \eqref{eq:fixedpoint2},
 the yellow dot denotes the repeller fixed point,
 and the black dot denotes the attractor (see the text).
 The two plots show different ranges in ${H}$ and $\dot\phi$. The left
 plot is a blowup of the right plot near the string-hole gas fixed point.
 Also in the left plot, the dashed gray curve depicts the curve $\dot\phi=dH$, along which the saddle point is unstable this time.}
\label{fig:ps2}
\end{figure*}

We now perform the same stability analysis as in the previous subsection, except starting with the EOM derived in Sec.~\ref{sec:alphaprimeMeissner}
for Meissner's action and setting $p=\sigma/2=0$. Here the fixed point is [recall Eq.~\eqref{eq:HsolSFSHgasw02}]
\begin{equation}
\label{eq:fixedpoint2}
 y^A_\star=(H_\star,\dot\phi_\star)=\frac{2}{\ell_\mathrm{s}}\sqrt{\frac{2\pi}{k(d-2)}}(1,d)~.
\end{equation}
As before, we put the set of differential equations in the form $\dot{y}^A=(\dot H,\ddot\phi)=\mathcal{C}^A(H,\dot\phi)$ and compute
the eigenvalues of the corresponding Jacobian matrix $J_A{}^B=\partial_A\mathcal{C}^B$ evaluated at the fixed point $y^A_\star$.
As a result, we find that there is one positive and one negative eigenvalue indicating that the fixed point is again a saddle point. 
This is confirmed by visual inspection of Fig.~\ref{fig:ps2} (see the left plot for a close-up;
Fig.~\ref{fig:ps2} is generated the same way as Fig.~\ref{fig:ps1}, in particular, setting $k=1$, $\ell_\mathrm{s}=1$, and $d=3$).
Contrary to the saddle point of the previous subsection, though, the saddle point here turns out to be unstable in the direction
of the string-hole gas constraint $H=\dot\phi/d$. Indeed, evaluating $\bm{D}_{\bm{u}}\mathcal{C}^A$ at $y^A_\star$ yields
two positive values. This is confirmed by looking at the direction of the flow along the dashed gray line in
the left plot of Fig.~\ref{fig:ps2}, which depicts the line $\dot\phi=dH$; the trajectories are moving away from the fixed point (in red).

Additional fixed points are found by numerically solving $\mathcal{C}^A(H,\dot\phi)=0$, and they are shown by the yellow and black dots in
Fig.~\ref{fig:ps2}. The black dot is an attractor and was also found in the context of vacuum pre-Big Bang cosmology
(see Ref.~\cite{Gasperini:2007zz}). However, in this case, one notices that the attractor fixed point is disconnected from
the trivial fixed point at the origin (see the right plot of Fig.~\ref{fig:ps2}),
which confirms the results of Refs.~\cite{Gasperini:2007zz,Brustein:1999yq}.
Furthermore, we find that the string-hole gas saddle point is also not connected to the string perturbative vacuum
as it was the case with GMV's action. In fact, trajectories that start near the origin tend to grow rapidly in $\dot\phi$,
while $H$ remains small, and go nowhere near the fixed points.

In summary, the string-hole gas fixed point, which is a solution of Meissner's action, shares several characteristics with
the solution of GMV's action: both are saddle points, disconnected from the string perturbative vacuum. However,
the trajectories in the vicinity of the saddle points behave very differently for both actions.
Indeed, the latter (GMV) is stable in the direction of the string-hole gas constraint $H=\dot\phi/d$,
but the former (Meissner) is unstable. Therefore, Meissner's action appears very unlikely to be the physical action
that can describe the evolution of a string-hole gas and of the universe at high energies.

Let us end by noting that, although the analysis outlined in this section focuses on the case $w=0$, we found that the qualitative results about the characterization and (in)stability of the fixed points are the same for any value of $w\in(-1,1/3)$. We do not include the quantitative details for a generic value of $w$ for the sake simplicity and readability.

\section{Conclusions and discussion}\label{sec:conclusions}

In this paper, we revisited the proposal that the stringy high-energy state of the Universe is a string-hole gas, i.e., a gas of black holes lying on the string-/black-hole correspondence curve. By analyzing its thermodynamic properties, we confirmed that a string-hole gas has the same EOS and entropy equation in the Einstein frame as a black-hole gas with $p=\rho$ and $S\sim\sqrt{EV/G}$. In the string frame, we found that a string-hole gas has vanishing pressure, and we derived the corresponding evolution to be given by $H=\dot\phi/d\sim\ell_\mathrm{s}^{-1}$. Our goal was then to find such a fixed point solution from the dynamical cosmological EOM of a string theory motivated action.
We studied the gravidilaton sector of the low-energy effective action of string theory and found that, to zeroth order in the $\alpha'$ expansion, there is no string-hole gas solution without adding a tuned negative cosmological constant. However, going to first order in $\alpha'$, we studied two different actions, and both yielded a natural string-hole gas solution.
Stability of those fixed point solutions was assessed by performing a phase space analysis. We found that both solutions are saddle points in $(H,\dot\phi)$ phase space, but the solution coming from the action of GMV \cite{Gasperini:1996fu} tends to be better behaved since it is stable in the direction of the string-hole gas constraint $H=\dot\phi/d$. The solution coming from the action of Meissner \cite{Meissner:1996sa} is unstable in the same direction and thus less appealing, even though it possesses the desired $O(d,d)$ symmetry of string cosmology to first order in $\alpha'$.
In summary, our results show that string theory consistently supports a string-hole gas phase of cosmological evolution, at least at the level of a gravidilaton effective action and minimally to first order in the $\alpha'$ expansion. Our stability analysis also indicates that a particular choice of action (GMV's action) is more appropriate at the level of our approximation.

We would like to point out some of the limitations of the current analysis. As mentioned before, the scale at which a string-hole gas forms and evolves is right at the limit of perturbative string theory in terms of the $\alpha'$ expansion.
Our analysis showed that one needs an action that is valid at least to first order in $\alpha'$, but one could seek for a yet higher-order action (e.g., to second order in $\alpha'$) or an exact conformal model (valid to all orders in $\alpha'$) for a more robust implementation.
Beforehand, it might be more straightforward to try to find a description of a string-hole gas such that its corresponding matter action has first-order $\alpha'$ corrections. Indeed, if first-order $\alpha'$ corrections are included in the gravity sector, they may as well be first-order $\alpha'$ corrections at the level of the matter action.
For example, higher energy-momentum tensor corrections in the matter sector have been considered in Ref.~\cite{Board:2017ign}
for Einstein gravity, but this has never been studied in the context of a string theory effective action or for any other theory beyond Einstein gravity. We note that such a possibility might also open the window to obtaining a nonsingular curvature bounce following the string-hole gas phase.

Another limitation comes from the fact that the current analysis was only performed within effective field theories of string theory and did not use perhaps the full `strength' of string theory. As future work, one could try to construct the proper matter action for string holes from first principles rather than using a thermodynamic approach. At the level of general relativity, there has been recent progress in describing a black-hole lattice in cosmology (see, in particular, Ref.~\cite{Clifton:2017hvg}
in a nonsingular bouncing cosmological background as well as, e.g., Ref.~\cite{BHlattice} and references therein), which may be viewed as an approximation of a black-hole gas. Similar ideas with the addition of appropriate stringy ingredients could be used to develop a nonperturbative action for a string-hole gas.

Let us also mention the fact that black holes in string theory may not be best described by the semiclassical picture used in this paper. The singularity at the center of black holes may be resolved in full string theory, and even the concept of a black-hole horizon may need to be revised. For instance, a stringy black hole might be better described by a `fuzzball' (see, e.g., Ref.~\cite{Fuzzballrefs} and references therein).
In that context, a black-hole gas may be realized as a set of intersecting brane states \cite{Masoumi:2014vpa}, which is related to the concept of fractional brane gas (see, e.g., Ref.~\cite{fractionalbranegas} and references therein).

Within the context of a string-hole gas as studied in this paper, we plan to extend the present work to determine what is the cosmological evolution subsequent to the string-hole gas phase and what the cosmological observable predictions intrinsic to the resulting very early Universe scenario are.
First, the goal is to determine how a string-hole gas phase can be connected to standard Big Bang cosmology starting with radiation-dominated expansion. A string-hole gas phase is not expected to be stable for an infinitely long period of time. The gas will ultimately (Hawking) evaporate into radiation \cite{Veneziano:2003sz},
a nonadiabatic process of entropy production that can be viewed as quantum particle creation in curved spacetime. Given that the string-hole gas is already saturating the appropriate entropy bound, the entropy release from the evaporation of the string holes cannot occur if the spacetime curvature remains constant or grows to a higher energy scale. Instead, the decay of the string holes must coincide with a (nonsingular) curvature bounce; in particular, the string- and Einstein-frames Hubble radii have to start growing. This would naturally coincide with the beginning of the expanding radiation-dominated phase of standard Big Bang cosmology.

Finding dynamics for the process of nonsingular curvature bounce shall be one of the key issues in follow-up work. Even though the actions studied in this paper contained higher-curvature corrections, they did not allow for nonsingular transitions from the string-hole gas phase to radiation expansion. Since the process of string-hole gas decay into radiation is quantum mechanical in nature, one may expect to find the desired dynamics from an action including quantum loop corrections. This is physically equivalent to taking into account the `backreaction' from particle production due to quantum fluctuations in curved spacetime \cite{Gasperini:2002bn}.
It is precisely this backreaction that might effectively violate the null energy condition, hence avoiding a Big Crunch singularity after the string-hole gas phase. Nonsingular bouncing backgrounds have already been found with string-theoretic loop corrections (see, e.g., Refs.~\cite{Brustein:1997cv,Cartier:1999vk,Tsujikawa:2002qc,Gasperini:2002bn,Gasperini:2007zz} and references therein), but never in the context of a string-hole gas phase. Loop corrections might not be the only way, though, to obtain a nonsingular bounce in string theory. Another possibility, for instance, would be to consider an S-brane, a stringy object that can prevent the Universe from reaching a Big Crunch (see Ref.~\cite{Sbranerefs}, also studied in Ref.~\cite{Brandenberger:2015nua}).

Finally, once a full very early Universe scenario has been developed at the background level, we shall be able to study the generation and evolution of the cosmological perturbations and determine what the observable predictions are. If fluctuations are seeded in the string-hole gas phase, one may find interesting results. On one hand, the quantum perturbations for a gas of black holes at the string scale may deviate considerably from the usual Bunch-Davies initial state. On the other hand, one shall not underestimate the effect of thermal fluctuations from the gas of string holes. Indeed, since the radius of the string holes equates the Hubble radius in the string frame, one may obtain holographic scaling of the specific heat capacity ($C_V\sim R^2$) on Hubble scales, similar to what is obtained from a string gas \cite{Brandenberger:2011et,Brandenberger:2008nx}. It shall be interesting to see what spectra of primordial perturbations result and how they differ from the results of string gas cosmology (see, e.g., Refs.~\cite{Brandenberger:2011et,Brandenberger:2008nx} and references therein), pre-Big Bang cosmology (see, e.g., Refs.~\cite{Gasperini:2002bn,Gasperini:2007zz,Gasperini:2007vw,Lidsey:1999mc} and references therein), and other very early Universe scenarios.

\begin{acknowledgments}
We would like to thank Samir Mathur for useful comments.
JQ acknowledges financial support from the Vanier Canada Graduate Scholarship
administered by the Natural Sciences and Engineering Research Council of Canada (NSERC).
RHB is supported in part by an NSERC Discovery grant and by the Canada Research Chair program.
MG is supported in part by the Istituto Nazionale di Fisica Nucleare (INFN) under the program Theoretical Astroparticle Physics.
\end{acknowledgments}

\appendix

\section{String-hole gas evolution in the Einstein frame}\label{sec:EFsol}

Given a consistent string-hole gas solution with $H=\dot\phi/d=\mathrm{constant}$ in the string frame,
one can derive the corresponding solution in the Einstein frame
by using the relation (see, e.g., Refs.~\cite{Gasperini:2002bn,Gasperini:2007zz})
\begin{equation}
 \tilde H=\left(H-\frac{\dot\phi}{d-1}\right)e^{\phi/(d-1)}~.
\end{equation}
In this Appendix, a tilde denotes an Einstein-frame quantity,
while no tilde means the string frame.
The constraint $H=\dot\phi/d$ thus implies
\begin{equation}
\label{eq:HESH}
 \tilde H=-\frac{H}{d-1}e^{\phi/(d-1)}~,
\end{equation}
so one notices that for a constant-Hubble expanding phase in the string frame ($H>0$),
the Einstein-frame Hubble parameter must be negative ($\tilde H<0$) and therefore contracting.

Let us recall that the Einstein-frame time is related to the string-frame time via (see, e.g., Ref.~\cite{Gasperini:2007zz})
\begin{equation}
\label{eq:EFtimedef}
 \mathrm{d}\tilde t=e^{-\phi/(d-1)}\mathrm{d}t~.
\end{equation}
Since $\dot\phi=dH=\mathrm{constant}$, where one now views $H=H_\star\sim\ell_\mathrm{s}^{-1}$ as one of the constant fixed-point solutions found, e.g.,
in Secs.~\ref{sec:alphaprimeGMV} or \ref{sec:alphaprimeMeissner}, one can write
\begin{equation}
 \phi(t)=dH(t-t_0)
\end{equation}
for $t\leq t_0$. The integration constant $t_0$ is set such that $\phi(t_0)=0$,
at which point $g_\mathrm{s}=e^{\phi/2}=1$, corresponding to strong coupling.
Thus, the evolution in the perturbative regime (where $g_\mathrm{s}\ll 1$) translates to $t\ll t_0$.
Upon integration of Eq.~\eqref{eq:EFtimedef}, one can then show that
\begin{equation}
\label{eq:tEFSH}
 \tilde t-\tilde t_0=-\frac{(d-1)}{dH}\left(e^{-\phi(t)/(d-1)}-1\right)~,
\end{equation}
for $\tilde t\leq\tilde t_0$, where $\tilde t_0$ in the Einstein-frame time equivalent to the string-frame time $t_0$.
Let us choose $\tilde t_0=-(d-1)/(dH)<0$, so then
\begin{equation}
\label{eq:ephiEFSH}
 e^{\phi/(d-1)}=\frac{d-1}{dH}\frac{1}{(-\tilde t)}~.
\end{equation}
Therefore, Eq.~\eqref{eq:HESH} becomes
\begin{equation}
\label{eq:HandasolEF}
 \tilde H(\tilde t)=-\frac{1}{d(-\tilde t)}~,
\end{equation}
which confirms $\tilde H<0$ since $\tilde t\leq\tilde t_0<0$. The above expression further implies
\begin{equation}
 \tilde a(\tilde t)\sim (-\tilde t)^{1/d}
\end{equation}
when integrating $\tilde H=\mathrm{d}\ln\tilde a/\mathrm{d}\tilde t$.
Combining with Eq.~\eqref{eq:ephiEFSH}, this implies $\tilde a\sim e^{-\phi/(d(d-1))}$,
which is in agreement with how one expects the Einstein-frame scale factor to behave for a string-hole gas
[recall Eq.~\eqref{eq:aevoEFSHgas}].

\end{document}